\documentclass[a4paper,UKenglish,cleveref,autoref,thm-restate]{lipics-v2021}

\pdfoutput=1

\usepackage{bm}
\usepackage{tikz}
	\usetikzlibrary{arrows}
	\usetikzlibrary{patterns}
\usepackage{tikz-cd}

\newcommand{\DS}{\mid\mkern3mu}				
\newcommand{\DP}{{.\kern5pt}}				
\newcommand{\DF}{\colon}				
\newcommand{\DE}{\mathrel{\mathop:}=}			
\newcommand{\ED}{=\mathrel{\mathop:}}			

\newcommand{\Enumref}[1]{\textbf{{#1}.}}		

\DeclareMathOperator{\supp}{supp}			

\title{Nawrotzki’s Algorithm for the Countable Splitting Lemma, Constructively}
\titlerunning{Constructive Nawrotzki algorithm}

\author{Ana Sokolova}{Department of Computer Sciences, University of Salzburg, Austria}{Ana.Sokolova@sbg.ac.at}{}{}
\author{Harald Woracek}{Institute of Analysis and Scientific Computing, TU Wien, Austria}{harald.woracek@tuwien.ac.at}{}{}
\authorrunning{A. Sokolova and H. Woracek}
\Copyright{Ana Sokolova and Harald Woracek}



\acknowledgements{We are indebted to Paul Levy for bringing this topic to us by asking us several years ago to figure out some
	details of Nawrotzki's algorithm, in particular its constructivity.}

\nolinenumbers
\hideLIPIcs

\begin{document}

\maketitle

\begin{abstract}
	We reprove the countable splitting lemma by adapting Nawrotzki's algorithm which produces a sequence that
	converges to a solution. Our algorithm combines Nawrotzki's approach with taking finite cuts. It is constructive in the
	sense that each term of the iteratively built approximating sequence as well as the error between the 
	approximants and the solution is computable with finitely many algebraic operations. 
\end{abstract}


%
%
%
\section{Explanation of what is going on ...}

Given a measure $\mu$ on a product space $\prod_{i\in I}X_i$, the \emph{$j$-th marginal} $\mu_j$ of $\mu$ is the push-forward of
$\mu$ under the $j$-th canonical projection $\pi_j\DF\prod_{i\in I}X_i\to X_j$. Explicitly, this is 
\[
	\mu_j(A)\DE\mu\big(\pi_j^{-1}(A)\big)
\]
for all $A\subseteq X_j$ with $\pi_j^{-1}(A)$ being measureable. 

In his fundamental paper \cite{strassen:1965} Strassen investigated the existence of measures on a product $X\times Y$ which
have prescribed marginals and satisfy additional constraints of a certain form. The result stated in \Cref{Y1} below is known as 
\emph{Strassen's theorem on stochastic domination}%
\footnote{%
	The result is a corollary of \cite[Theorem~11]{strassen:1965}. Curiously, it is not even explicitly stated in
	Strassen's paper, but only mentioned in one sentence.
	}.
The stated variant is taken from \cite[Corollary~7]{skala:1993}%
\footnote{%
	A different proof can be found in \cite{kellerer:1984}.
	}. 
To formulate it, we need some notation.
\begin{itemize}
\item Let $X$ be a Hausdorff space, and let $\preccurlyeq$ be a partial order on $X$ which is closed as a subset of $X\times X$. 
	A subset $A\subseteq X$ is \emph{upward closed w.r.t.\ $\preccurlyeq$}, if 
	\[
		\forall x\in X,y\in A\DP y\preccurlyeq x\Rightarrow x\in A
		.
	\]
\item For two positive Borel measures $\mu,\nu$ on $X$ we write \emph{$\mu\preceq\nu$}, if for all upward closed Borel sets
	$A\subseteq X$ it holds that $\mu(A)\leq\nu(A)$. 
\end{itemize}

\begin{theorem}
\label{Y1}
	Let $X$ be a Hausdorff space, let $\preccurlyeq$ be a closed partial order on $X$, and let $\mu$ and $\nu$ be two
	probability (Borel-) measures on $X$. If $\mu\preceq\nu$, then there exists a probability (Borel-) measure 
	$\Lambda$ on $X\times X$ which has the marginals $\mu$ and $\nu$, and whose support is contained in $\preccurlyeq$. 
\end{theorem}

\noindent
An important particular case of \Cref{Y1} is when the base space $X$ is finite or countable with the discrete topology. 

Over the years this result was established on different levels of generality; some papers are 
\cite{aniello.wright:2000}, 
\cite{berti.pratelli.rigo.spizzichino:2015},
\cite{friedland.ge.zhi:2020},
\cite{gaffke.rueschendorf:1984},
\cite{kawabe:2000},
\cite{koenig:2012},
\cite{kamae.krengel.brien:1977}.
Some predecessors of Strassen's work are \cite{kellerer:1961,nawrotzki:1962}. 

\Cref{Y1} plays an important role in probability theory and has applications in various areas. For example, it prominently occurs
in finance mathematics, e.g.\ \cite{beiglboeck.juillet:2016,davis.hobson:2007}, or in computer science, e.g.\ 
\cite{hsu:2017,jones:1989,jung.tix:1998}. 

The proof of \Cref{Y1} relies in general on a rather heavy analytic machinery, in particular, on theorems exploiting 
compactness properties. If $X$ is finite, a required solution $\Lambda$ can -- naturally -- be found by an algorithm which
terminates after finitely many steps. This fact can be based on various reasoning. For example on elementary manipulations with
inequalities, as e.g.\ in \cite[\S3]{kellerer:1961}, or combinatorial results like the max-flow min-cut theorem or the subforest
lemma, as e.g.\ in \cite{koperberg:2016} or \cite[Theorem~4.10]{jones:1989}. 

In the present exposition we deal with the countable discrete case. Our aim is to give a recursive algorithm which produces 
a sequence $(\Delta_N)_{N\in\mathbb N}$ of (discrete) probability measures on $X\times X$ such that 
\begin{enumerate}
\item each term of the sequence is computable from the inital data $\mu,\nu$ with a finite number of algebraic operations;
\item the sequence $(\Delta_N)_{N\in\mathbb N}$ converges to a solution $\Lambda$ in the $\ell^1$-norm on $X\times X$, in
	particular it converges pointwise;
\item the speed of pointwise convergence can be controlled in a computable way. 
\end{enumerate}
To explain our contribution, it is worthwhile to revisit the presently availabe proofs for the countable discrete case. 
First, specialising the general proof(s) of \Cref{Y1} obviously does not lead to an algorithm, since tools like e.g.\ the
Banach-Alaoglu Theorem are used. More interesting are the arguments given in the papers of Kellerer \cite[\S4]{kellerer:1961}
and Nawrotzki \cite{nawrotzki:1962}. Both are inconstructive, but for different reasons. 
\begin{itemize}
\item Kellerer's approach is to reduce to the finite cases. Given $\mu,\nu$ on a countable set, he produces appropriately
	cut-off data $\mu_N,\nu_N$, $N\in\mathbb N$, and solves the problem for those. 
	This gives a measure $\Lambda_N$ on $X$, which
	solves the problem up to the index $N$. Each measure $\Lambda_N$ can be computed in finitely many steps. 
	Sending the cut-off point $N$ to infinity leads to existence of a solution for the full data $\mu,\nu$. 
	The masses of the measures $\Lambda_N$ may oscillate, and therefore the sequence $(\Lambda_N)_{N\in\mathbb N}$ need not be
	convergent. However, each accumulation point of the sequence $(\Lambda_N)_{N\in\mathbb N}$ will be a solution. 

	What makes the method inconstructive is that accumulation points \emph{exist by compactness} (in this case applied in 
	the form of the Heine-Borel Theorem). 
\item Nawrotzki's approach is to produce a sequence $(\Lambda_N)_{N\in\mathbb N}$, which does not necessarily solve the problem on
	any finite section, but still converges to a solution. His construction ensures that the masses of the measures 
	$\Lambda_N$ are nonincreasing on points of the diagonal and nondecreasing off the diagonal. This ensures that passing to
	subsequences is not necessary. 

	What makes the method inconstructive is that defining the measures $\Lambda_N$ requires to evaluate 
	\emph{sums of infinite series} and \emph{infima of infinite sets} of real numbers.
\end{itemize}
Our idea to produce $(\Delta_N)_{N\in\mathbb N}$ with \Enumref{1}--\Enumref{3} above, is to combine the approaches: 
we apply Nawrotzki's algorithm to appropriately truncated sequences to ensure computability, and control the error which is 
made by passing to cut-off's to ensure convergence.

\section{Nawrotzki's algorithm}

In \cite{nawrotzki:1962}, which preceeds the work of Strassen, Nawrotzki proved a discrete version of Strassen's theorem. In our
present language his result reads as follows.

\begin{theorem}
\label{Y50}
	Let $\mu=(\mu_n)_{n\in\mathbb N}$ and $\nu=(\nu_n)_{n\in\mathbb N}$ be sequences of real numbers, such that 
	\begin{equation}
	\label{Y51}
		\forall n\in\mathbb N\DP \mu_n\geq 0\wedge\nu_n\geq 0
		\quad\text{and}\quad
		\sum_{n\in\mathbb N}\mu_n=\sum_{n\in\mathbb N}\nu_n=1
		,
	\end{equation}
	Moreover, let $\preccurlyeq$ be a partial order on $\mathbb N$. 
	
	If it holds that 
	\begin{equation}
	\label{Y53}
		\forall R\subseteq\mathbb N\text{ upwards closed w.r.t.\ $\preccurlyeq$}\DP
		\sum_{n\in R}\mu_n\leq\sum_{n\in R}\nu_n
		,
	\end{equation}
	then there exists an infinite matrix $\Lambda=(\lambda_{n,m})_{n,m\in\mathbb N}$ of real numbers, such that 
	\begin{align}
		& \forall n,m\in\mathbb N\DP \lambda_{n,m}\geq 0
		\quad\text{and}\quad
		\sum_{n,m\in\mathbb N}\lambda_{n,m}=1
		,
		\label{Y54}
		\\
		& \forall n,m\in\mathbb N\DP \lambda_{n,m}\neq 0\Rightarrow n\preccurlyeq m
		,
		\label{Y56}
		\\[2mm]
		& \forall n\in\mathbb N\DP \sum_{m\in\mathbb N}\lambda_{n,m}=\mu_n
		,
		\label{Y57}
		\\
		& \forall m\in\mathbb N\DP \sum_{n\in\mathbb N}\lambda_{n,m}=\nu_m
		.
		\label{Y58}
	\end{align}
\end{theorem}

\noindent
In this section we present Nawrotzki's argument in a structured way including all details. This provides an in-depth
understanding of his work, and this is necessary to make appropriate adaption to the algorithm later on (in Section~3). 

\begin{remark}
\label{Y99}
	Before we dive into the formulas and proofs, which are a bit technical and lengthy, let us give an intuition for what is
	going to happen. 

	Assume we are given data $\mu_n,\nu_m$ satisfying \cref{Y51,Y53} and a (probably bad) approximation of
	a solution $\lambda_{n,m}$ that satisfies \cref{Y54,Y56}, as well as \cref{Y57}. 
	Note that achieving correctness of one marginal, i.e.\ satisfying \cref{Y57}, is very easy; 
	for example already the diagonal matrix with $\mu_n$'s on the diagonal will satisfy this. 

	If the column sums do not give the correct results as required by \cref{Y58}, it must be that some of them are larger 
	than the target value and some of them are smaller since the total sum is always $1$. Now we want to modify the values 
	$\lambda_{n,m}$ to improve the approximation, i.e., make the error in \cref{Y58} smaller while retaining all other
	properties. Most importantly, we have to ensure that \cref{Y53}, also known as \emph{stochastic dominance}, is inherited.
	In addition, we want to make the modification in such a way that:
	\begin{enumerate}
	\item At each place $(n,m)$ entries change monotonically when repeating the step in the algorithm. This is achieved by
		having diagonal entries nonincreasing and off-diagonal entries nondecreasing. This will guarantee existence of a
		limit. 
	\item Make sure that the pattern of which column sums are too large and	which are too small is inherited with exception
		that some column sums may become correct. This will guarantee that the algorithm can proceed appropriately. 
	\end{enumerate}
	The algorithm proceeds in steps. In each step exactly two values of the matrix change: one at the diagonal at position
	$(n,n)$ and another in the same row at position $(n,m)$ such that \cref{Y58} fails for $n$ and $m$, as pictured below.
	The new values are $\lambda'_{n,n}=\lambda_{n,n}-\alpha$ and $\lambda'_{n,m}=\lambda_{n,m}+\alpha$, where $\alpha$ is
	chosen such that still $\lambda'_{*,n}\geq\nu_n$, $\lambda'_{*,m}\leq\nu_m$. 

	In the picture, filled circles indicate those points where our approximation has nonzero entries, 
	circled dots mark the changes made by one step of the algorithm, and $\alpha>0$ is the correction term whose 
	exact definition (see \cref{Y65}) is taylor made so that the above explained requirements are met.
	
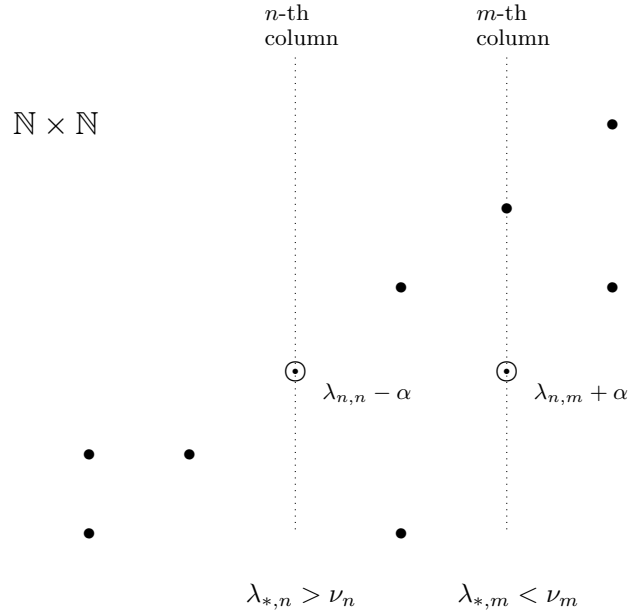
\begin{center}
	\begin{tikzcd}
		&& \parbox{0mm}{\hspace*{-4mm}\parbox{10mm}{\footnotesize{$n$-th column}}} && 
		\parbox{0mm}{\hspace*{-4mm}\parbox{10mm}{\footnotesize{$m$-th column}}} &
		\\
		\parbox{0mm}{\hspace*{-10mm}\parbox{20mm}{\Large$\mathbb N\times\mathbb N$}} && 
		\phantom{} && \phantom{} & \bullet
		\\
		&&&& \bullet &
		\\
		&&& \bullet && \bullet
		\\
		&& 
		{\bm\odot}\parbox{0mm}{\hspace*{2mm}\vspace*{-6mm}\parbox{20mm}{\footnotesize$\lambda_{n,n}-\alpha$}} && 
		{\bm\odot}\parbox{0mm}{\hspace*{2mm}\vspace*{-6mm}\parbox{20mm}{\footnotesize$\lambda_{n,m}+\alpha$}} &
		\\
		\bullet & \bullet &&&&
		\\
		\bullet && \phantom{} \ar[-,dotted]{uuuuuu} & \bullet & \phantom{} \ar[-,dotted]{uuuuuu} &
		\\[-3mm]
		&& \parbox{1mm}{\hspace*{-6mm}\parbox{20mm}{$\lambda_{*,n}>\nu_n$}} && 
		\parbox{1mm}{\hspace*{-6mm}\parbox{20mm}{$\lambda_{*,m}<\nu_m$}} &
	\end{tikzcd}
	\end{center}
\end{remark}

\noindent
The next result, \Cref{Y60}, is the first crucial ingredient to Nawrotzki's algorithm 
(out of two; the second is \Cref{Y31} further below). It will ensure that in the limit a solution is obtained.
To formulate it, we need additional notation.

\begin{definition}
\label{Y59}
	Let $\preccurlyeq$ be a partial order on $\mathbb N$. 
	For each $(n,m)\in\mathbb N\times\mathbb N$ with $n\prec m$, we denote 
	\[
		\mathcal R_{n,m}\DE \big\{R\subseteq\mathbb N\DS n\notin R,m\in R,
		R\text{ upward closed w.r.t.\ $\preccurlyeq$}\big\}
		.
	\]
\end{definition}

\noindent
Note that $\mathcal R_{n,m}$ is always nonempty. For example, we have 
\[
	\{l\in\mathbb N\DS m\preccurlyeq l\}\in\mathcal R_{n,m}
	.
\]

\begin{proposition}
\label{Y60}
	Assume that $\mu$, $\nu$, and $\preccurlyeq$, satisfy \cref{Y51} and \cref{Y53}. 
	If for each pair $(n,m)\in\mathbb N\times\mathbb N$ with $n\prec m$ at least one of
	\begin{align}
		& \mu_n\leq\nu_n
		,
		\label{Y61}
		\\
		& \mu_m\geq\nu_m
		,
		\label{Y62}
		\\
		& \inf_{R\in\mathcal R_{n,m}}\sum_{l\in R}(\nu_l-\mu_l)=0
		,
		\label{Y63}
	\end{align}
	holds, then $\mu=\nu$. 
\end{proposition}

\noindent
Note here that all series in \cref{Y63} converge absolutely and that by \cref{Y53} the infimum in \cref{Y63} is nonnegative.
Moreover, in an algorithm acting as explained in \Cref{Y99} above (and defined in precise mathematical terms in \Cref{Y65}
below), using $\mathcal R_{n,m}$ instead of all upwards closed sets is sufficient to retain \cref{Y53}. This is because for
upwards closed sets which are not in $\mathcal R_{n,m}$, \cref{Y53} is trivially inherited. 

In the proof of \Cref{Y60}, we use the following simple fact.

\begin{lemma}
\label{Y64}
	Assume that $\mu$, $\nu$, and $\preccurlyeq$, satisfy \cref{Y51} and \cref{Y53}. 
	Further, let $R_1,R_2,\ldots$ be a (finite or infinite) sequence of upward closed (w.r.t.\ $\preccurlyeq$) subsets of
	$\mathbb N$, and set 
	\[
		R\DE \bigcup_k R_k
		.
	\]
	Then $R$ is upward closed, and 
	\[
		\sum_{l\in R}(\nu_l-\mu_l)\leq\sum_k\sum_{l\in R_k}(\nu_l-\mu_l)
		.
	\]
\end{lemma}
\begin{proof}
	By absolute convergence we may rearrange the sum on the left side without changing its value. Now write $R$ as the
	disjoint union 
	\[
		R=\dot\bigcup_k R'_k
	\]
	where 
	\[
		R'_k\DE R_k\setminus\bigcup_{j<k}R_j
		.
	\]
	Then 
	\[
		\sum_{l\in R}(\nu_l-\mu_l)=\sum_k\sum_{l\in R'_k}(\nu_l-\mu_l)
		.
	\]
	For each $k$ we have 
	\[
		\sum_{l\in R_k}(\nu_l-\mu_l)=\sum_{l\in R'_k}(\nu_l-\mu_l)+
		\sum_{R_k\cap\bigcup_{j<k}R_j}(\nu_l-\mu_l)
		.
	\]
	The set $R_k\cap\bigcup_{j<k}R_j$ is upward closed, and hence the second summand on the right side is nonnegative. This
	shows that 
	\[
		\sum_{l\in R'_k}(\nu_l-\mu_l)\leq\sum_{l\in R_k}(\nu_l-\mu_l)
	\]
	for all $k$.
\end{proof}

\begin{proof}[Proof of \Cref{Y60}]
	It is enough to show that $\mu_n\leq\nu_n$ for all $n\in\mathbb N$. Assume towards a contradiction that there exists 
	$n\in\mathbb N$ with $\mu_n>\nu_n$, and fix one with this property. Moreover, choose $\epsilon>0$ small enough, say, 
	\[
		\epsilon\DE\frac 13(\mu_n-\nu_n)
		.
	\]
	By the assumption of the proposition we know that for each $m\in\mathbb N$ with $m\succ n$ at least one of 
	\begin{itemize}
	\item $\mu_m\geq\nu_m$,
	\item $\inf_{R\in\mathcal R_{n,m}}\sum_{l\in R}(\nu_l-\mu_l)=0$,
	\end{itemize}
	must hold. 

	Consider the set where the second case takes place
	\[
		H\DE\Big\{m\in\mathbb N\DS n\prec m,\inf_{R\in\mathcal R_{n,m}}\sum_{l\in R}(\nu_l-\mu_l)=0\Big\}
		.
	\]
	If $H=\emptyset$, it is easy to reach a contradiction. Namely, if $\mu_m\geq\nu_m$ for all $m\succ n$, then 
	\[
		\sum_{m\succcurlyeq n}\mu_m>\sum_{m\succcurlyeq n}\nu_m
		,
	\]
	and this contradicts \cref{Y53}. 

	If $H\neq\emptyset$, we argue as follows. For each $m\in H$ choose $R_m\in\mathcal R_{n,m}$, such that 
	\[
		\sum_{l\in R_m}(\nu_l-\mu_l)\leq\frac\epsilon{2^m}
		,
	\]
	and set $R\DE\bigcup_{m\in H}R_m$. Then $H\subseteq R$, $n\notin R$, and 
	\[
		\sum_{l\in R}(\nu_l-\mu_l)\leq\sum_{m\in H}\sum_{l\in R_m}(\nu_l-\mu_l)
		\leq\sum_{m\in H}\frac\epsilon{2^m}\leq 2\epsilon
		.
	\]
	Consider the upward closed set 
	\[
		R'\DE R\cup\{l\in\mathbb N\DS n\prec l\}
		.
	\]
	If $l\in R'\setminus R$, then $n\prec l$ and $l\notin H$. Thus we must have $\mu_l\geq\nu_l$. From this we see that 
	\[
		0\leq\sum_{l\in R'}(\nu_l-\mu_l)=
		\sum_{l\in R}(\nu_l-\mu_l)+\sum_{l\in R'\setminus R}(\nu_l-\mu_l)
		\leq\sum_{l\in R}(\nu_l-\mu_l)\leq 2\epsilon
		.
	\]
	The set $R'\cup\{n\}$ is also upward closed. Using the above estimate, and recalling that $n\notin R'$, we reach the
	contradiction 
	\[
		0\leq\sum_{l\in R'\cup\{n\}}(\nu_l-\mu_l)=
		\sum_{l\in R'}(\nu_l-\mu_l)+(\nu_n-\mu_n)
		\leq 2\epsilon+(\nu_n-\mu_n)=\frac 13(\nu_n-\mu_n)<0
		.
	\]
\end{proof}

\noindent
Nawrotzki's algorithm for the proof of \Cref{Y50} proceed in three steps:
\begin{enumerate}
\item Start with the diagonal matrix built from $\mu$.
\item Iteratively modify this matrix in such a way, that the set of all points $(n,m)$ where all of \cref{Y61}--\cref{Y63} fail
	(for certain modified sequences), gets smaller in each step.
\item Pass to the limit, so to reach a situation where \Cref{Y60} applies.
\end{enumerate}
The single steps of the recursive process \Enumref{2} are realised by maps which act on 
$\ell^1(\mathbb N\times\mathbb N)$. To define those 
maps, we first introduce an abbreviation for row- and column sums of a matrix. 
Given $\Lambda=(\lambda_{n,m})_{n,m\in\mathbb N}\in\ell^1(\mathbb N\times\mathbb N)$, we denote 
\[
	\lambda_{*,m}\DE\sum_{n\in\mathbb N}\lambda_{n,m},\quad 
	\lambda_{n,*}\DE\sum_{m\in\mathbb N}\lambda_{n,m}
	.
\]
Note that these series converge absolutely since $\Lambda\in\ell^1(\mathbb N\times\mathbb N)$. 

\begin{definition}
\label{Y65}
	Let $\nu=(\nu_n)_{n\in\mathbb N}\in\ell^1(\mathbb N)$ and $(n,m)\in\mathbb N\times\mathbb N$. We define maps 
	\[
		\alpha_{n,m}^\nu\DF\ell^1(\mathbb N\times\mathbb N)\to[0,\infty)
		,\quad
		\Phi_{n,m}^\nu\DF\ell^1(\mathbb N\times\mathbb N)\to\ell^1(\mathbb N\times\mathbb N)
		.
	\]
	\begin{itemize}
	\item For $\Lambda\in\ell^1(\mathbb N\times\mathbb N)$ set 
		\[
			\alpha_{n,m}^\nu(\Lambda)\DE\min\Big\{\lambda_{*,n}-\nu_n,\nu_m-\lambda_{*,m},
			\inf_{R\in\mathcal R_{n,m}}\sum_{l\in R}(\nu_l-\lambda_{*,l})\Big\}
			,
		\]
		if $n\preccurlyeq m$ and this minimum is positive, and set $\alpha_{n,m}^\nu\DE 0$ otherwise. 
	\item For $\Lambda\in\ell^1(\mathbb N\times\mathbb N)$ let $\Phi_{n,m}^\nu(\Lambda)$ be the matrix with the entries 
		\[
			\Big[\Phi_{n,m}^\nu\Big]_{l,k}(\Lambda)\DE
			\begin{cases}
				\lambda_{l,k}-\alpha_{n,m}^\nu(\Lambda) &\text{if}\ (l,k)=(n,n)
				,
				\\
				\lambda_{l,k}+\alpha_{n,m}^\nu(\Lambda) &\text{if}\ (l,k)=(n,m)
				,
				\\
				\lambda_{l,k} &\text{otherwise}
				.
			\end{cases}
		\]
	\end{itemize}
\end{definition}

\noindent
Note that $\Phi_{n,m}^\nu$ is well-defined, since $\alpha_{n,m}^\nu\neq 0$ implies that $n\neq m$, and since it is obvious that 
$\Phi_{n,m}^\nu(\Lambda)$ is again summable. 

Let us collect some more obvious properties of the transformations $\Phi_{n,m}^\nu$. 

\begin{remark}
\label{Y66}
	For each $\nu\in\ell^1(\mathbb N)$ and $(n,m)\in\mathbb N\times\mathbb N$, the following statements hold. 
	\begin{enumerate}
	\item ${\displaystyle
		\supp\Phi_{n,m}^\nu(\Lambda)\subseteq\big(\supp\Lambda\big)\cup\{(n,n),(n,m)\}
		}$%
		,
	\item ${\displaystyle
		\forall l\in\mathbb N\DP \Big[\Phi_{n,m}^\nu(\Lambda)\Big]_{l,*}=\lambda_{l,*}
		}$%
		,
	\item ${\displaystyle
		\forall l\in\mathbb N\DP \Big[\Phi_{n,m}^\nu(\Lambda)\Big]_{*,l}=
		\begin{cases}
			\lambda_{*,l}-\alpha_{n,m}^\nu(\Lambda) &\text{if}\ l=n
			,
			\\
			\lambda_{*,l}+\alpha_{n,m}^\nu(\Lambda) &\text{if}\ l=m
			,
			\\
			\lambda_{*,l} &\text{otherwise}
			.
		\end{cases}
		}$%
	\end{enumerate}
\end{remark}

\noindent
Having $\alpha_{n,m}^\nu(\Lambda)=0$ just means that at the point $(n,m)$ one of \cref{Y61}--\cref{Y63} holds for the sequences
$(\lambda_{*,n})_{n\in\mathbb N}$ and $(\nu_n)_{n\in\mathbb N}$. Moreover, in this case, $\Phi_{n,m}^\nu$ does not change $\Lambda$. 
We are interested to see what happens if $\alpha_{n,m}^\nu(\Lambda)>0$.

\begin{definition}
\label{Y67}
	Let $\nu\in\ell^1(\mathbb N)$ and $\Lambda\in\ell^1(\mathbb N\times\mathbb N)$. Then we set 
	\[
		S(\Lambda)\DE\Big\{(n,m)\in\mathbb N\times\mathbb N\DS \alpha_{n,m}^\nu(\Lambda)>0\Big\}
		.
	\]
	Moreover, we denote by $\pi_1(S(\Lambda))$ and $\pi_2(S(\Lambda))$ the projections of $S(\Lambda)$ onto the first and
	second, respectively, component. 
\end{definition}

\noindent
To avoid bulky notation, we do not explicitly notate the dependency on $\nu$.
Moreover, observe that $S(\Lambda)$ is contained in $\preccurlyeq$ and does not intersect the diagonal, in fact, 
\[
	 \pi_1(S(\Lambda))\cap\pi_2(S(\Lambda))=\emptyset
	 .
\]
In the next proposition we show that $\Phi_{n,m}^\nu$ preserves several relevant properties and indeed shrinks the set
$S(\Lambda)$.

\begin{proposition}
\label{Y31}
	Let $\nu=(\nu_n)_{n\in\mathbb N}\in\ell^1(\mathbb N)$, $\Lambda\in\ell^1(\mathbb N\times\mathbb N)$, and assume that 
	\begin{align}
		& \forall n,m\in\mathbb N\DP \lambda_{n,m}\geq 0
		\quad\text{and}\quad 
		\sum_{n,m\in\mathbb N}\lambda_{n,m}=1
		,
		\label{Y32}
		\\
		& \forall n\in\pi_1(S(\Lambda))\DP \lambda_{*,n}=\lambda_{n,n}
		.
		\label{Y35}
		\\[3mm]
		& \forall R\subseteq\mathbb N\text{ upward closed w.r.t.\ $\preccurlyeq$}\DP
		\sum_{l\in R}\lambda_{*,l}\leq\sum_{l\in R}\nu_l
		,
		\label{Y34}
	\end{align}
	Further, let $(n',m')\in\mathbb N\times\mathbb N$, and assume that $\alpha_{n',m'}^\nu(\Lambda)>0$.
	Then 
	\begin{enumerate}
	\item $\Phi_{n',m'}^\nu(\Lambda)$ satisfies \cref{Y32}, \cref{Y35}, and \cref{Y34},
	\item ${\displaystyle
		S\big(\Phi_{n',m'}^\nu(\Lambda)\big)\subseteq S(\Lambda)\setminus\{(n',m')\}
		}$%
		.
	\end{enumerate}
\end{proposition}
\begin{proof}
	To shorten notation, we write
	\[
		\Lambda'=(\lambda'_{n,m})_{n,m\in\mathbb N}\DE\Phi_{n',m'}^\nu(\Lambda)
		.
	\]
	We start with showing that $\Lambda'$ satisfies \cref{Y32} and \cref{Y34}.
	Let $(n,m)\neq (n',n')$. Then $\lambda'_{n,m}\geq\lambda_{n,m}$ and hence is nonnegative. 
	For $(n,m)=(n',n')$ we use \eqref{Y35} to obtain 
	\[
		\lambda'_{n',n'}=\lambda_{n',n'}-\alpha_{n',m'}^\nu(\Lambda)=\lambda_{*,n'}-\alpha_{n',m'}^\nu(\Lambda)
		\geq\nu_{n'}\geq 0
		.
	\]
	Obviously, applying $\Phi_{n',m'}^\nu$ does not change the total sums of the entries of a matrix.
	Thus
	\[
		\sum_{n,m\in\mathbb N}\lambda'_{n,m}=\sum_{n,m\in\mathbb N}\lambda_{n,m}=1
		.
	\]
	We see that \cref{Y32} holds. 

	Let $R\subseteq\mathbb N$ be upward closed. If $R\notin\mathcal R_{n',m'}$, then 
	\[
		\sum_{l\in R}\lambda'_{*,l}\leq\sum_{l\in R}\lambda_{*,l}\leq\sum_{l\in R}\nu_l
		.
	\]
	Next, for $R\in\mathcal R_{n',m'}$
	\begin{equation}
	\label{Y71}
		\sum_{l\in R}\lambda'_{*,l}=\sum_{l\in R}\lambda_{*,l}+\alpha_{n',m'}^\nu(\Lambda)
		,
	\end{equation}
	and from this we find 
	\[
		\sum_{l\in R}\lambda'_{*,l}=\sum_{l\in R}\lambda_{*,l}+\alpha_{n',m'}^\nu(\Lambda)\leq
		\sum_{l\in R}\lambda_{*,l}+\sum_{l\in R}(\nu_n-\lambda_{*,l})=\sum_{l\in R}\nu_l
		.
	\]
	Thus \cref{Y34} holds. 

	Now we come to the proof of \Enumref{2}. This is the major part of the argument.

	In the first step we show that $(n',m')\notin S(\Lambda')$. We make a case distinction according to 
	which term is the minimum in the definition of $\alpha_{n',m'}^\nu(\Lambda)$. 
	\begin{itemize}
	\item Case $\alpha_{n',m'}^\nu(\Lambda)=\lambda_{*,n'}-\nu_{n'}$:
		\\[1mm]
		Then $\lambda'_{*,n'}=\nu_{n'}$, and hence $n'\notin\pi_1(S(\Lambda'))$. In particular, 
		$(n',m')\notin S(\Lambda')$. 
	\item Case $\alpha_{n',m'}^\nu(\Lambda)=\nu_{m'}-\lambda_{*,n'}$:
		\\[1mm]
		Then $\lambda'_{*,m'}=\nu_{m'}$, and hence $m'\notin\pi_2(S(\Lambda'))$. In particular, 
		$(n',m')\notin S(\Lambda')$. 
	\item Case $\alpha_{n',m'}^\nu(\Lambda)=\inf_{\mathcal R_{n',m'}}\sum_{l\in R}(\nu_l-\lambda_{*,l})$:
		\\[1mm]
		Recalling \cref{Y71}, we find 
		\[
			\inf_{R\in\mathcal R_{n',m'}}\sum_{l\in R}(\nu_l-\lambda'_{*,l})=
			\inf_{R\in\mathcal R_{n',m'}}\sum_{l\in R}\Big[(\nu_l-\lambda_{*,l})-\alpha_{n',m'}^\nu(\Lambda)\Big]
			=0
			.
		\]
		Thus also in this case $(n',m')\notin S(\Lambda')$. 
	\end{itemize}
	In the second step, we show that $S(\Lambda')\subseteq S(\Lambda)$. Assume towards a contradiction that 
	$(n,m)\in S(\Lambda')\setminus S(\Lambda)$. Explicitly this means that 
	\begin{align*}
		n\prec m\ &\wedge\ \lambda'_{*,n}>\nu_n\ \wedge\ \lambda'_{*,m}<\nu_m\ \wedge\ 
		\inf_{R\in\mathcal R_{n,m}}\sum_{l\in R}(\nu_l-\lambda'_{*,l})>0
		\\
		&\wedge\ \Big[\ \lambda_{*,n}\leq\nu_n\ \vee\ \lambda_{*,m}\geq\nu_m\ \vee\ 
		\inf_{R\in\mathcal R_{n,m}}\sum_{l\in R}(\nu_l-\lambda_{*,l})=0\ \Big]
	\end{align*}
	We distinguish cases according to the disjunction in the square bracket. 
	\begin{itemize}
	\item Case $\lambda_{*,n}\leq\nu_n$:
		\\[1mm]
		The sum of the $n$-th column increases, and thus we must have $n=m'$. This implies 
		\[
			\lambda'_{*,n}=\lambda'_{*,m'}=\lambda_{*,m'}+\alpha_{n',m'}^\nu(\Lambda)\leq\nu_{m'}=\nu_n
			,
		\]
		which contradicts the second term in the conjunction. 
	\item Case $\lambda_{*,m}\geq\nu_m$:
		\\[1mm]
		The sum of the $m$-th column decreases, and thus we must have $m=n'$. This implies 
		\[
			\lambda'_{*,m}=\lambda'_{*,n'}=\lambda_{*,n'}-\alpha_{n',m'}^\nu(\Lambda)\geq\nu_{n'}=\nu_m
			,
		\]
		which contradicts the third term in the conjunction. 
	\item Case $\inf_{R\in\mathcal R_{n,m}}\sum_{l\in R}(\nu_l-\lambda_{*,l})=0$:
		\\[1mm]
		Choose $R'\in\mathcal R_{n,m}$ such that 
		\[
			\sum_{l\in R'}(\nu_l-\lambda_{*,l})<\inf_{R\in\mathcal R_{n,m}}\sum_{l\in R}(\nu_l-\lambda'_{*,l})
			.
		\]
		Then, in particular, the value of the sum over all $l\in R'$ decreases, and we must have $n'\in R'$ and 
		$m'\notin R'$. Since $R'$ is upward closed and $n'\prec m'$, this is a contradiction. 
	\end{itemize}
	The proof of \Enumref{2} is complete. 

	It remains to deduce \cref{Y35}. Let $n\in\pi_1(S(\Lambda'))$. Then also $n\in\pi_1(S(\Lambda))$, and therefore 
	$n\neq m'$ and $\lambda_{*,n}=\lambda_{n,n}$. From the first property we obtain that the $n$-th column is 
	modified at most at its diagonal entry, and now the second implies that $\lambda'_{*,n}=\lambda'_{n,n}$.
\end{proof}

\noindent
Next, we investigate iterative application of maps $\Phi_{n,m}^\nu$. Start with $\nu\in\ell^1(\mathbb N)$, 
$\Lambda^{(0)}\in\ell^1(\mathbb N\times\mathbb N)$, and a sequence $((n_k,m_k))_{k\geq 1}$ of points in $\mathbb N\times\mathbb N$. 
From this data, we built the sequence $(\Lambda^{(k)})_{k\in\mathbb N}$ where 
\begin{equation}
\label{Y68}
	\Lambda^{(k)}\DE\Big[\Phi_{n_k,m_k}^\nu\circ\cdots\circ\Phi_{n_1,m_1}^\nu\Big]\big(\Lambda^{(0)}\big)
	.
\end{equation}
It turns out that, in the situation of \Cref{Y50}, sequences of this form converge. 
In fact, they do so because of a very simple reason, namely, monotonicity.

\begin{lemma}
\label{Y70}
	Let $(\Lambda^{(k)})_{k\in\mathbb N}$ be a sequence in $\ell^1(\mathbb N\times\mathbb N)$, such that 
	\[
		\sup_{k\in\mathbb N}\|\Lambda^{(k)}\|_1<\infty,\quad \forall n,m,k\in\mathbb N\DP \lambda_{n,m}^{(k)}\geq 0
		,
	\]
	and that there exists a partition $\mathbb N\times\mathbb N=A\dot\cup B$ such that $(\lambda_{n,m}^{(k)})_{k\in\mathbb N}$ is
	nondecreasing for all $(n,m)\in A$ and nonincreasing for all $(n,m)\in B$. 

	Then the limit $\Lambda\DE\lim_{k\to\infty}\Lambda^{(k)}$ exists in the $\ell^1$-norm.
\end{lemma}
\begin{proof}
	Each of the sequences $(\lambda_{n,m}^{(k)})_{k\in\mathbb N}$ is monotone and bounded, hence convergent. 
	Denote $\lambda_{n,m}\DE\lim_{k\to\infty}\lambda_{n,m}^{(k)}$. We have to show that the pointwise limit 
	$\Lambda=(\lambda_{n,m})_{n,m\in\mathbb N}$ is actually attained in the $\ell^1$-norm. To this end we split the
	corresponding sum according to the given partition.

	For each $(n,m)\in A$ the sequence $(\lambda_{n,m}^{(k)})_{k\in\mathbb N}$ is nondecreasing, and hence the 
	monotone convergence theorem yields 
	\[
		\sum_{(n,m)\in A}\lambda_{n,m}=\lim_{k\to\infty}\sum_{(n,m)\in A}\lambda_{n,m}^{(k)}
		\leq\sup_{k\in\mathbb N}\|\Lambda^{(k)}\|_1<\infty
		.
	\]
	Since $\lambda_{n,m}\geq\lambda_{n,m}-\lambda_{n,m}^{(k)}\geq 0$, we may now refer to the bounded convergence theorem to
	obtain that 
	\[
		\lim_{k\to\infty}\sum_{(n,m)\in A}\big|\lambda_{n,m}^{(k)}-\lambda_{n,m}\big|=0
		.
	\]
	For each $(n,m)\in B$ and $k\in\mathbb N$ we have 
	\[
		\lambda_{n,m}^{(0)}\geq\lambda_{n,m}^{(k)}\geq\lambda_{n,m}^{(k)}-\lambda_{n,m}\geq 0
		.
	\]
	Since $\sum_{(n,m)\in B}\lambda_{n,m}^{(0)}<\infty$, the bounded convergence theorem applies, and we find that 
	\[
		\lim_{k\to\infty}\sum_{(n,m)\in B}\big|\lambda_{n,m}^{(k)}-\lambda_{n,m}\big|=0
		.
	\]
\end{proof}

\begin{corollary}
\label{Y41}
	Assume that $\Lambda^{(0)}$ satisfies \cref{Y32} and \cref{Y35}, let $((n_k,m_k))_{k\geq 1}$ be any 
	sequence, and let $(\Lambda^{(k)})_{k\in\mathbb N}$ be defined by \cref{Y68}. 
	Then the limit 
	\[
		\Lambda\DE\lim_{k\to\infty}\Lambda^{(k)}
	\]
	exists w.r.t.\ the $\ell^1$-norm.
\end{corollary}
\begin{proof}
	Since $\alpha_{n,m}^\nu(\Lambda)$ is always nonnegative, a partition of $\mathbb N\times\mathbb N$ required to apply 
	\Cref{Y70} is obtained by taking the diagonal as the set $A$.
\end{proof}

\noindent
Now we show that, when passing to a limit, the set $S(\Lambda)$ can be controlled. 

\begin{lemma}
\label{Y39}
	Let $(\Lambda^{(k)})_{k\in\mathbb N}$ be a sequence in $\ell^1(\mathbb N\times\mathbb N)$ which converges in the 
	$\ell^1$-norm, and denote $\Lambda\DE\lim_{k\to\infty}\Lambda^{(k)}$. Then 
	\[
		S(\Lambda)\subseteq\bigcup_{N\in\mathbb N}\bigcap_{k\geq N} S(\Lambda^{(k)})
		.
	\]
\end{lemma}
\begin{proof}
	Let $(n,m)\in S(\Lambda)$, and set $\epsilon\DE\frac 12\alpha_{n,m}^\nu(\Lambda)$. 
	Choose $N\in\mathbb N$ such that 
	\[
		\forall k\geq N\DP \|\Lambda^{(k)}-\Lambda\|_1\leq\epsilon
		.
	\]
	Then for all $k\geq N$ 
	\[
		\lambda_{*,n}^{(k)}\geq\lambda_{*,n}-\epsilon\geq\nu_n
		,\quad
		\lambda_{*,m}^{(k)}\leq\lambda_{*,m}+\epsilon\leq\nu_m
		,
	\]
	and for all $R\in\mathcal R_{n,m}$ 
	\[
		\sum_{l\in R}\big(\nu_l-\lambda_{*,l}^{(k)}\big)\geq\sum_{l\in R}(\nu_l-\lambda_{*,l})-\epsilon
		\geq\epsilon>0
	\]
	Thus $(n,m)\in S(\Lambda^{(k)})$.
\end{proof}

\noindent
We have collected all the neccessary tools needed for the proof of \Cref{Y50}.

\begin{proof}[Proof of \Cref{Y50}]
	Let $\mu$, $\nu$, and $\preccurlyeq$, be given, and assume that \cref{Y51} and \cref{Y53} hold. 

	Let $\Lambda^{(0)}=(\lambda^{(0)}_{n,m})_{n,m\in\mathbb N}$ be the diagonal matrix built from $\mu$, i.e., 
	\begin{equation}
	\label{Y28}
		\lambda_{n,m}^{(0)}\DE
		\begin{cases}
			\mu_n &\text{if}\ n=m
			,
			\\
			0 &\text{otherwise}
			.
		\end{cases}
	\end{equation}
	Choose a sequence of points $((n_k,m_k))_{k\geq 1}$ in $\mathbb N\times\mathbb N$ which covers $\prec$. 
	For example, every enumeration of $\mathbb N\times\mathbb N$ certainly has this property. 
	Now define $\Lambda^{(k)}$ by \cref{Y68} using this sequence. 

	By \Cref{Y31}, each $\Lambda^{(k)}$ satisfies \cref{Y32}, \cref{Y35}, and \cref{Y34}. Moreover, 
	\[
		S(\Lambda^{(k)})\subseteq S(\Lambda^{(0)})\setminus\big\{(n_1,m_1),\ldots,(n_k,m_k)\big\}
		.
	\]
	The limit 
	\[
		\Lambda=(\lambda_{n,m})_{n,m\in\mathbb N}\DE\lim_{k\to\infty}\Lambda^{(k)}
	\]
	exists in the $\ell^1$-norm by \Cref{Y41}, and $S(\Lambda)=\emptyset$ by \Cref{Y39}.

	Clearly, \cref{Y54}--\cref{Y57} hold for $\Lambda$. By virtue of \Cref{Y31}, we may apply \Cref{Y60} with the sequences
	$(\lambda_{*,n})_{n\in\mathbb N}$ and $(\nu_n)_{n\in\mathbb N}$, and obtain that also \cref{Y58} holds. 
\end{proof}

\noindent
We refer to the procedure carried out in this proof as \emph{Nawrotzki's algorithm} being \emph{performed along} the sequence 
$((n_k,m_k))_{k\geq 1}$. 

\begin{remark}
\label{Y3}
	For later use, we observe the following fact. Let $(\Lambda^{(k)})_{k\in\mathbb N}$ be a 
	sequence produced by an application
	of Nawrotzki's algorithm. Then off-diagonal elements $\lambda^{(k)}_{n,m}$ change their value at most once when $k$ runs
	through $\mathbb N$. Namely, only when $(n,m)=(n_k,m_k)$ and it happens that $\alpha^\nu_{n,m}(\Lambda^{(k-1)})>0$. 
\end{remark}

\section{A constructive variant of the algorithm}

Nawrotzki's proof of \Cref{Y50} is inconstructive for the following reason:
\begin{itemize}
\item The set $\mathcal R_{n,m}$ is in general infinite, and its elements themselves are in general infinite. 
\end{itemize}
Because of this, computing the numbers $\alpha_{n,m}^\nu$ requires to evaluate the sum of infinite series and an infimum of an
infinite set. Hence, it is not possible to compute any term of the sequence $(\Lambda^{(k)})_{k\in\mathbb N}$, which converges to a
solution matrix $\Lambda$, with a finite number of algebraic operations. 

Our aim is to give a proof of \Cref{Y50} which is more constructive in the following sense.

\begin{theorem}
\label{Y12}
	Let $\mu,\nu,\preccurlyeq$ be given such that \cref{Y51} and \cref{Y53} hold. Then there exists a sequence 
	$(\Delta^{(k)})_{k\in\mathbb N}$ of matrices in $\ell^1(\mathbb N\times\mathbb N)$ with the following properties. 
	\begin{enumerate}
	\item Each $\Delta^{(k)}$ can be computed from the given data $\mu$ and $\nu$ by a finite number of algebraic operations. 
	\item The limit $\Delta\DE\lim_{k\to\infty}\Delta^{(k)}$ exists in the $\ell^1$-norm and satisfies 
		\cref{Y54}--\cref{Y58}. 
	\end{enumerate}
	As usual we use the notation $\Delta^{(k)}=(\delta^{(k)}_{n,m})_{n,m\in\mathbb N}$ and 
	$\Delta=(\delta_{n,m})_{n,m\in\mathbb N}$.
	\begin{enumerate}
	\setcounter{enumi}{2}
	\item For each fixed $(n,m)\in\mathbb N\times\mathbb N$ with $n\prec m$, and for each $\epsilon>0$, a number $k_0$ 
		with the property that 
		\[
			\forall k\geq k_0\DP |\delta_{n,m}^{(k)}-\delta_{n,m}|\leq\epsilon
		\]
		can be computed from the given data $\mu$ and $\nu$ by a finite number of algebraic operations
	\end{enumerate}
\end{theorem}

\noindent
While the speed of pointwise convergence is controlled by the assertion in item \Enumref{3} (even in a constructive way), we have
no control of the speed of $\ell^1$-convergence.

The idea to prove this theorem is the simplest possible: we consider cut-off data $\mu_N$, $\nu_N$ instead of $\mu$, $\nu$, apply 
Nawrotzki's algorithm to the truncated data, and then send the cut-off point to infinity. 
Realising this idea, however, requires some work. 

We start with discussing convergence matters. The error when using cut-off's instead of the full data can be controlled using the
following general perturbation lemma. 

\begin{lemma}
\label{Y7}
	Let $\nu,\tilde\nu\in\ell^1(\mathbb N)$, $\Lambda,\tilde\Lambda\in\ell^1(\mathbb N\times\mathbb N)$, and $(n,m)\in\mathbb N\times\mathbb N$. 
	Then 
	\begin{equation}
	\label{Y9}
		\big|\alpha^\nu_{n,m}(\Lambda)-\alpha^{\tilde\nu}_{n,m}(\tilde\Lambda)\big|\leq
		\|\Lambda-\tilde\Lambda\|_1+\|\nu-\tilde\nu\|_1
		.
	\end{equation}
\end{lemma}
\begin{proof}
	We have 
	\begin{multline*}
		\big|(\lambda_{*,n}-\nu_n)-(\tilde\lambda_{*,n}-\tilde\nu_n)\big|
		\\
		\leq\sum_{l\in\mathbb N}|\lambda_{l,n}-\tilde\lambda_{l,n}|+|\nu_n-\tilde\nu_n|
		\leq\|\Lambda-\tilde\Lambda\|_1+\|\nu-\tilde\nu\|_1
		,
	\end{multline*}
	and in the same way 
	\begin{multline*}
		\big|(\lambda_{*,m}-\nu_m)-(\tilde\lambda_{*,m}-\tilde\nu_m)\big|
		\\
		\leq\sum_{l\in\mathbb N}|\lambda_{l,m}-\tilde\lambda_{l,m}|+|\nu_m-\tilde\nu_m|
		\leq\|\Lambda-\tilde\Lambda\|_1+\|\nu-\tilde\nu\|_1
		.
	\end{multline*}
	Next let $R\subseteq\mathbb N$. Then 
	\begin{multline*}
		\Big|\sum_{l\in R}(\nu_l-\lambda_{*,l})-\sum_{l\in R}(\tilde\nu_l-\tilde\lambda_{*,l})\Big|\leq
		\\
		\leq\sum_{l\in R}\sum_{k\in\mathbb N}|\lambda_{k,l}-\tilde\lambda_{k,l}|+\sum_{l\in R}|\nu_l-\tilde\nu_l|
		\leq \|\Lambda-\tilde\Lambda\|_1+\|\nu-\tilde\nu\|_1
		.
	\end{multline*}
	It follows that 
	\begin{align*}
		\bigg|
		\inf\Big(\{\lambda_{*,n}- &\, \nu_n,\nu_m-\lambda_{*,m}\}\cup
		\Big\{\sum_{l\in R}(\nu_l-\lambda_{*,l})\DS R\in\mathcal R_{n,m}\Big\}\Big)
		\\
		&\, -
		\inf\Big(\{\tilde\lambda_{*,n}-\tilde\nu_n,\tilde\nu_m-\tilde\lambda_{*,m}\}\cup
		\Big\{\sum_{l\in R}(\tilde\nu_l-\tilde\lambda_{*,l})\DS R\in\mathcal R_{n,m}\Big\}\Big)
		\bigg|
		\\
		&\, \leq \|\Lambda-\tilde\Lambda\|_1+\|\nu-\tilde\nu\|_1
	\end{align*}
	This is \cref{Y9} if $n\preccurlyeq m$. Otherwise $\alpha^\nu_{n,m}=\alpha_{n,m}^{\tilde\nu}(\tilde\Lambda)=0$, 
	and the required estimate holds trivially.
\end{proof}

\begin{corollary}
\label{Y10}
	Let $\nu,\tilde\nu\in\ell^1(\mathbb N)$, $\Lambda,\tilde\Lambda\in\ell^1(\mathbb N\times\mathbb N)$, and 
	$((n_k,m_k))_{k\geq 1}$ be a sequence in $\mathbb N\times\mathbb N$. Let $(\Lambda^{(k)})_{k\in\mathbb N}$ and 
	$(\tilde\Lambda^{(k)})_{k\in\mathbb N}$ be the sequences defined by \cref{Y68} starting from 
	$\Lambda^{(0)}\DE\Lambda$ and $\tilde\Lambda^{(0)}\DE\tilde\Lambda$, respectively. Moreover, set 
	\[
		\epsilon\DE\|\Lambda-\tilde\Lambda\|_1+\|\nu-\tilde\nu\|_1
		.
	\]
	Then 
	\[
		\forall k\in\mathbb N\DP \|\Lambda^{(k)}-\tilde\Lambda^{(k)}\|_1+\|\nu-\tilde\nu\|_1\leq 3^k\epsilon
		.
	\]
\end{corollary}
\begin{proof}
	For $k=0$ this is the definition of $\epsilon$. Then proceed inductively based on the estimate 
	\[
		\big\|\Phi_{n,m}^\nu(\Lambda)-\Phi_{n,m}^{\tilde\nu}(\tilde\Lambda)\big\|_1
		\leq\|\Lambda-\tilde\Lambda\|_1+2|\alpha_{n,m}^\nu(\Lambda)-\alpha_{n,m}^{\tilde\nu}(\tilde\Lambda)|
		,
	\]
	which holds for all $\nu,\tilde\nu,\Lambda,\tilde\Lambda,n,m$.
\end{proof}

\noindent
Now we turn to computability matters. To settle these, we need one more notation.

\begin{definition}
\label{Y13}
	Let $L\subseteq\mathbb N$, and $n,m\in L$ with $n\prec m$. Then we set 
	\[
		\mathcal R^L_{n,m}\DE\big\{R\subseteq L\DS n\notin R,m\in R,
		\forall k\in R,l\in L\DP k\preccurlyeq l\Rightarrow l\in R\big\}
		.
	\]
\end{definition}

\begin{lemma}
\label{Y8}
	Let $\nu\in\ell^1(\mathbb N)$, $\Lambda\in\ell^1(\mathbb N\times\mathbb N)$, let $L\subseteq\mathbb N$, and assume that 
	\begin{equation}
	\label{Y18}
		\supp\nu\subseteq L,\quad \supp\Lambda\subseteq L\times L
		.
	\end{equation}
	Then 
	\begin{align}
		\label{Y14}
		& \forall (n,m)\notin L\times L\DP \alpha_{n,m}^\nu(\Lambda)=0
		,
		\\
		\label{Y15}
		& \forall (n,m)\in\mathbb N\times\mathbb N\DP \supp\Phi_{n,m}^\nu(\Lambda)\subseteq L\times L
		,
		\\
		\label{Y16}
		& \forall n,m\in L,n\prec m\DP 
		\inf_{R\in\mathcal R_{n,m}}\sum_{l\in R}(\nu_l-\lambda_{*,l})
		=\inf_{R\in\mathcal R^L_{n,m}}\sum_{l\in R}(\nu_l-\lambda_{*,l})
		.
	\end{align}
\end{lemma}
\begin{proof}
	The assumption on the supports of $\nu$ and $\Lambda$ shows that 
	\[
		\forall n\notin L\DP \nu_n=\lambda_{*,n}=0
		.
	\]
	From this \cref{Y14}, and in turn also \cref{Y15}, follows. Moreover, for every subset $R\subseteq\mathbb N$
	\[
		\sum_{l\in R}(\nu_l-\lambda_{*,l})=\sum_{l\in R\cap L}(\nu_l-\lambda_{*,l})
		.
	\]
	To establish \cref{Y16}, we show that for all $n,m\in L$ with $n\prec m$
	\[
		\mathcal R_{n,m}^L=\{R\cap L\DS R\in\mathcal R_{n,m}\}
		.
	\]
	The inclusion ``$\supseteq$'' is clear. For the reverse inclusion observe that, for each $R\in\mathcal R_{n,m}^L$, the set 
	\[
		R'\DE\big\{l\in\mathbb N\DS \exists k\in R\DP k\preccurlyeq l\big\}
	\]
	belongs to $\mathcal R_{n,m}$ and $R'\cap L=R$.
\end{proof}

\begin{corollary}
\label{Y17}
	Let $\nu\in\ell^1(\mathbb N)$ and $\Lambda\in\ell^1(\mathbb N\times\mathbb N)$ be finitely supported. Then 
	\begin{enumerate}
	\item for each $n\in\mathbb N$ the number $\lambda_{*,n}$ is a finite sum, and 
	\item for each $(n,m)\in\mathbb N\times\mathbb N$ the infimum in the definition of $\alpha_{n,m}^\nu(\Lambda)$ is 
		the minimum of a finite number of finite sums. 
	\end{enumerate}
\end{corollary}
\begin{proof}
	We can choose a finite set $L\subseteq\mathbb N$ such that \cref{Y18} holds. 
	Then each set $\mathcal R_{n,m}^L$, and also each of
	its elements, is finite. 
\end{proof}

\begin{proof}[Proof of \Cref{Y12}]
	Consider truncated data: for $N\in\mathbb N$, let $\mu_N=(\mu_{N;n})_{n\in\mathbb N}$ and 
	$\nu_N=(\nu_{N;n})_{n\in\mathbb N}$ be defined by 
	\[
		\mu_{N;n}\DE
		\begin{cases}
			\mu_n &\text{if}\ n<N
			,
			\\
			1-\sum_{l<N}\mu_l &\text{if}\ n=N
			,
			\\
			0 &\text{if}\ n>N
			,
		\end{cases}
		\qquad
		\nu_{N;n}\DE
		\begin{cases}
			\nu_n &\text{if}\ n<N
			,
			\\
			1-\sum_{l<N}\nu_l &\text{if}\ n=N
			,
			\\
			0 &\text{if}\ n>N
			.
		\end{cases}
	\]
	We execute Nawrotzki's algorithm with the data $\mu_N,\nu_N$ along the enumeration $((n_k,m_k))_{k\geq 1}$ of 
	$\mathbb N\times\mathbb N$ which is defined by running through the scheme 
	
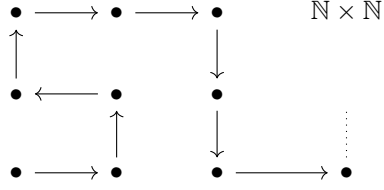
\begin{center}
	\begin{tikzcd}
		\bullet \ar{r} & \bullet \ar{r} & \bullet \ar{d} & \mathbb N\times\mathbb N
		\\
		\bullet \ar{u} & \bullet \ar{l} & \bullet \ar{d} & \phantom{}
		\\
		\bullet \ar{r} & \bullet \ar{u} & \bullet \ar{r} & \bullet \ar[dotted,-]{u}
	\end{tikzcd}
	\end{center}
	and dropping all points $(n,m)$ which do not satisfy $n\prec m$. 

	This provides us with sequences $(\Lambda_N^{(k)})_{k\in\mathbb N}$, $N\in\mathbb N$. 
	According to \Cref{Y8} and \Cref{Y17}, we have 
	\[
		\supp\Lambda_N^{(k)}\subseteq\{0,\ldots,N\}\times\{0,\ldots,N\}
		,
	\]
	and each $\Lambda_N^{(k)}$ can be computed by a finite number of algebraic operations. 

	Let $(\Lambda^{(k)})_{k\in\mathbb N}$ be the sequence obtained by running Nawrotzki's algorithm along the same sequence
	$((n_k,m_k))_{k\geq 1}$ but starting with the full data $\mu,\nu$. We have 
	\[
		\|\Lambda^{(0)}-\Lambda_N^{(0)}\|_1=2\sum_{n>N}\mu_n
		,\quad 
		\|\nu-\nu\|_1=2\sum_{n>N}\nu_n
		,
	\]
	and hence 
	\[
		\|\Lambda^{(0)}-\Lambda_N^{(0)}\|_1+\|\nu-\nu\|_1=
		2\sum_{n>N}(\mu_n+\nu_n)=2\Big(2-\sum_{n\leq N}(\mu_n+\nu_n)\Big)\ED\epsilon_N
		.
	\]
	\Cref{Y10} applies and leads to the basic estimate 
	\begin{equation}
	\label{Y11}
		\forall k\in\mathbb N,N\in\mathbb N\DP \|\Lambda^{(k)}-\Lambda_N^{(k)}\|_1+\|\nu-\nu\|_1\leq 3^k\epsilon_N
		.
	\end{equation}
	The next step is to define a sequence $(\Delta_k)_{k\in\mathbb N}$. This is done as follows: given $k\in N$, choose 
	$N_k\in\mathbb N$ with 
	\[
		\epsilon_{N_k}\leq\frac 1{k\cdot 3^k}
		,
	\]
	and set $\Delta_k\DE\Lambda_{N_k}^{(k)}$. 

	The number $N_k$ can be found in finitely many steps by summing up beginning sections of $\mu$ and $\nu$. Together with
	what we already observed above, thus, each $\Delta_k$ can be computed in finitely many steps. 

	We know that the limit $\Lambda\DE\lim_{k\to\infty}\Lambda^{(k)}$ exists in the $\ell^1$-norm and satisfies 
	\cref{Y54} -- \cref{Y58}. The basic estimate \cref{Y11} yields 
	\[
		\|\Lambda^{(k)}-\Delta^{(k)}\|_1\leq\frac 1k
		,
	\]
	and we see that also $\lim_{k\to\infty}\Delta^{(k)}=\Lambda$ in the $\ell^1$-norm.

	Let $(n,m)\in\mathbb N\times\mathbb N$ with $n\prec m$ and $\epsilon>0$ be given. 
	Define $k_0\in\mathbb N$ as the least integer larger or equal to 
	\[
		\max\Big\{\frac 1\epsilon,\big(\max\{n,m\}\big)^2\Big\}
		.
	\]
	Then $(n,m)\in\{(n_1,m_1),\ldots,(n_{k_0},m_{k_0})\}$ and for all $k\geq k_0$
	\[
		\|\Lambda^{(k)}-\Delta^{(k)}\|_1\leq\epsilon
		.
	\]
	Now recall \Cref{Y3}: the entry $\lambda_{n,m}^{(k)}$ is constant for $k\geq k_0$. This implies that, 
	forall $k\geq k_0$,
	\[
		|\lambda_{n,m}-\delta_{n,m}^{(k)}|=|\lambda_{n,m}^{(k)}-\delta_{n,m}^{(k)}|\leq
		\|\Lambda^{(k)}-\Delta^{(k)}\|_1\leq\epsilon
		.
	\]
	The proof of \Cref{Y12} is complete.
\end{proof}


%
%
%
\end{document}